\documentclass{article}
\usepackage{amsmath}
\usepackage{wrapfig}
\usepackage{pgfplots}
\usepackage{enumerate}
\usepackage{amssymb}
\usepackage{algorithm}
\input{Qcircuit}
\usepackage[noend]{algpseudocode}
\usepackage{amsmath}
\usepackage[justification=centering]{caption}
\numberwithin{equation}{section}
\usepackage{authblk}
\usepackage{hyperref}
\usepackage[misc]{ifsym}
\usepackage{amsthm}
\usepackage{cite}

\theoremstyle{definition}
\newtheorem{definition}{Definition}[section]

\usepackage{amsmath,amssymb}
\makeatletter
\newsavebox\myboxA
\newsavebox\myboxB
\newlength\mylenA

\newcommand*\xoverline[2][0.75]{%
    \sbox{\myboxA}{$\m@th#2$}%
    \setbox\myboxB\null
    \ht\myboxB=\ht\myboxA%
    \dp\myboxB=\dp\myboxA%
    \wd\myboxB=#1\wd\myboxA
    \sbox\myboxB{$\m@th\overline{\copy\myboxB}$}
    \setlength\mylenA{\the\wd\myboxA}
    \addtolength\mylenA{-\the\wd\myboxB}%
    \ifdim\wd\myboxB<\wd\myboxA%
       \rlap{\hskip 0.5\mylenA\usebox\myboxB}{\usebox\myboxA}%
    \else
        \hskip -0.5\mylenA\rlap{\usebox\myboxA}{\hskip 0.5\mylenA\usebox\myboxB}%
    \fi}
\makeatother

\providecommand{\keywords}[1]{\textbf{\textit{Keywords: }} #1}
\def\PACSname{\textbf{PACS}\enspace}
\def\PACS#1{\par\addvspace\medskipamount{\rightskip=0pt plus1cm
\def\and{\ifhmode\unskip\nobreak\fi\ $\cdot$
}\noindent\PACSname\ignorespaces#1\par}}

\title{A Quantum Algorithm for Testing Junta Variables and Learning Boolean Functions via Entanglement Measure}
\date{}

\author[1]{
  Khaled El-Wazan \footnote{khaled\_elwazan@alex-sci.edu.eg (\Letter)}
}

\author[1,2]{
  Ahmed Younes \thanks{ayounes@alexu.edu.eg}
}

\author[1]{
 	S. B. Doma \thanks{sbdoma@yahoo.com}
}

\affil[1]{Department of Mathematics and Computer Science, Faculty of Science, Alexandria University, Egypt}
\affil[2]{School of Computer Science, University of Birmingham, Birmingham, B15 2TT, United Kingdom}

\begin{document}

\maketitle

\begin{abstract}

Given a black-box representing an unknown Boolean function $f$ of $n$ variables, in this paper we propose a fast quantum algorithm to test whether or not a certain variable in the function $f$ is a junta variable. The proposed algorithm creates entanglement between the variable under test and an auxiliary qubit, where the  entanglement is measured using concurrence measure to decide if the variable is junta. The paper shows applications to the proposed algorithm in learning and categorization of Boolean functions.

\end{abstract}

\section{Introduction}
\label{introduction}
Quantum computers \cite{Feynman1986,deutsch,loyed} are powerful probabilistic computational devices which harness quantum phenomena like entanglement and superposition, to create problem solving paradigms more powerful than that of classical counterpart \cite{buhrman}. For instance, L. Grover introduced an optimal quantum search algorithm \cite{grover,grover-optimal} to search for an item among list of unstructured items in quadratic speed-up compared to classical search algorithms. Boyer \textit{el al.} later generalized Grover search algorithm \cite{boyer} to search for multiple items in a list of unstructured items, also in quadratic speed-up.  P. Shor presented a quantum algorithm \cite{shor} to find the prime factors of integer numbers in polynomial time.

Einstein \textit{et al.} were the first to characterize entanglement \cite{EPR}, and it was later discovered the potentials of such strange correlation as a useful resource for computations and information processing \cite{neilson-chuang}, with crucial utilization in many applications, for instance, quantum search algorithm \cite{younes-miller,younes} with reliable behavior, satellite-based quantum key distribution \cite{quantum-satellite,quantum-satellite-exp}, and quantum internet \cite{quantum-internet}. For those reasons, the need to create entanglement drives many experimental research groups to reach that goal, and recently there has been significant advances \cite{cloud,ion-trapped,photons}. Still in many applications that utilize entanglement, it is necessary useful to detect such correlation and quantify it theoretically and experimentally. Many entanglement detection methods for bipartite and multipartite  systems have been proposed \cite{li-zhao,zhou,zhang,regula-adesso}, based on popular entanglement measures \cite{e-detection}, for instance, entanglement of formation, distance measurement and concurrence. 


During the last two decades, the computation industry has been facing a real challenge due to the increasing amount of data \cite{makinsey-report} that needs to be handled and processed for certain purposes, and the increasing demand of the market to introduce faster computation devices that can cope up with the increasing amount of data as convenient as possible. One of the purposes for processing such data is the requirement to decide whether a given data has a certain property or not. But considering the enormous size of data, a classical algorithm testing all entries in that data for containing a certain property might not be practical or even feasible. 
One of the novel methods in property testing algorithms is quantum testing of classical properties, in which a quantum algorithm is utilized by harnessing quantum phenomena to introduce quantum speed-ups compared to classical property testing algorithms. The unique nature of quantum testing algorithms allows global extraction of properties and information about the input and the nature of the data itself by querying  all the possible entries at the same time.

One of the important classical properties that gained attention lately in the last decade \cite{survey} is junta property testing. Let $f$ be a Boolean function with $n$ inputs that maps a binary vector to either $0$ or $1$, \textit{i.e.} $f:\{0,1\}^n\rightarrow \{0,1\}$, a junta variable $x_i$ is an input to the function $f$ which the function does not depend on to perform such mapping. The junta property is considered an important typical problem, for instance, in machine learning techniques where it is vital to discriminate between relevant and irrelevant features to the learning process \cite{blum1,blum2}.

The first quantum $k$-junta $\epsilon$-tester \cite{atici-servedo} for Boolean functions and based on Fourier Sampling \cite{Bernstein-Vazirani}, due to At\i c\i~and Servedio in 2008, uses $\mathcal{O}(k/\epsilon)$ queries to the tested function, given the number of variables $k$ that the function depends on in advance. In addition, a quantum learning algorithm was introduced for $k$-junta Boolean functions to accuracy $\epsilon$ that requires $\mathcal{O}(k/\epsilon \log{k})$ quantum examples and $\mathcal{O}(2^k \log{\epsilon^{-1}})$ random examples.

In 2015, Li and Yang introduced a quantum algorithm \cite{yang} based on Bernstein-Vazirani algorithm \cite{Bernstein-Vazirani} to measure the influence of variables in Boolean functions. They also discussed other probabilistic quantum  learning algorithms for quadratic and cubic functions of simple forms.

Ambainis \textit{et al.} presented a quantum algorithm \cite{Ambainis} for testing junta functions, which is based on an algorithm that solves the group testing problem \cite{Sterrett}. Their quantum junta tester is found to require $\mathcal{\tilde{O}}(\sqrt{k/\epsilon})$ oracle calls.

In 2017, we introduced a quantum algorithm \cite{elwazan} to test junta variables in Boolean functions, based on a quantum search algorithm \cite{younes-miller} with reliable behavior. 
This algorithm requires two oracle calls to the black-box representing the Boolean  function to create a superposition with certain properties for junta variable testing. This algorithm requires $\mathcal{O}(\sqrt{2^n})$ oracle calls.

The aim of this paper is to present a quantum algorithm for testing variables in Boolean functions for the junta property.  The proposed algorithm transforms the problem of testing junta property in variables to measuring entanglement between qubits, using a reliable entanglement measure. The proposed algorithm uses a quantum operator $U_\lambda$ that creates entanglement between the tested variable $x_i$ for the junta property and an auxiliary qubit, if and only if the tested qubit is in superposition. The proposed algorithm is further used to find variables residing with the tested variable $x_i$ in function definition, and to discriminate between constant, balanced or Boolean functions of other form.

This paper is organized as follows: Section ~\ref{preliminaries} introduces the basic necessary concepts. 
Section ~\ref{proposed-operator} describes the proposed operator $U_\lambda$. 
Section ~\ref{proposed-algorithm} will cover the proposed algorithm along with the analysis. 
Section ~\ref{learn} introduces applications of the proposed algorithm for learning Boolean functions, followed by a conclusion in Section  ~\ref{conclusion}.

\section{Preliminaries}
\label{preliminaries}

\subsection{Problem Definition}
\label{problem-den}
One of the important classical properties that might exist in a data set is the junta property. 
We define the problem of finding whether a variable is relevant or irrelevant to a given Boolean function as follows:
\begin{definition}
Given a data set $\Lambda$ 	of all possible vectors of size $N=2^n$, and a Boolean function $f(x_0,x_1,\cdots,x_{n-1})$ which is used to map each vector $x\in\Lambda$ to either $\{0,1\}$, it is required to find whether the variable $x_i$ in the Boolean function contributes to mapping the input $x\in\Lambda$ to either $0$ or $1$.
\end{definition}

For $k$ inputs such that $k\leq n$, we say that a Boolean function $f$ is a $k$-junta function, if this Boolean function depends on at most $k$ out of its $n$ input variables.

For a linear Boolean function represented in positive polarity Reed-Muller \cite{reed-muller}:

\begin{align}
f(x_0,x_1,\cdots,x_{n-1})&=c_0x_0\oplus c_1x_1\oplus \cdots \oplus c_{n-1}x_{n-1}\nonumber\\
&=\bigoplus_{i=0}^{n-1}c_ix_i
\end{align}
where $\oplus$ is the addition modulo $2$ and the coefficient $c_i$ dictates whether the variable $x_i$ resides in the function definition or not, Bernstein-Vazirani algorithm \cite{Bernstein-Vazirani} can be used to find the exact solution of this form of Boolean functions in $\mathcal{O}(1)$, from which we can identify the junta variables. In this paper, we are interested to test if a given variable is junta in a general form Boolean function \cite{reed-muller}:

\begin{align}
f(x_0,x_1,...,x_i,...,x_{n-1})=\bigoplus_{q=0}^{N-1}c_qP_q \textit{ , }
\label{generalformfun}
\end{align}
where
\begin{align*}
P_q&: \emph{product term} \\
c_q&=
\begin{cases}
0:\emph{ product term does not exist}\\
1: \emph{product term exists}
\end{cases}.
\end{align*}

\subsection{Measures of Entanglement}
\label{measures-of-entanglement}
\begin{definition}
Given a quantum system $\ket{\varphi}$ of $2$ qubits, the pure bipartite state $\ket{\varphi}$ is called separable or a product state, if we can find states $\ket{\phi_A}$ and $\ket{\phi_B}$ in Hilbert space $\mathcal{H}$ of dimension $2$ such that 
\begin{align}
\ket{\varphi}=\ket{\phi_A}\otimes \ket{\phi_B},
\end{align}
otherwise, the state $\ket{\varphi}$ is called entangled.

\end{definition}

Concurrence measure \cite{hill} is considered one of the most popular measures of entanglement quantification of bipartite systems, and can be defined as follows \cite{walborn}:
\begin{equation}
C(\ket{\varphi})=\vert \bra{\varphi}(\sigma_y\otimes \sigma_y)\ket{\varphi^\dagger} \vert
\label{eq:concurrence-ge}
\end{equation} 
where $\sigma_y=-\imath\ket{1}\bra{0}+\imath\ket{0}\bra{1}$, such that $\imath=\sqrt{-1}$.

%

\section{The Proposed Operator}
\label{proposed-operator}

\begin{definition}
Let's assume having a quantum register of $n$ qubits and given two qubits indexed $i$ and $j$ in this quantum register. An entanglement measure device $D_{i,j}$ is a device which measures the entanglement between those designated qubits. This defined device can be illustrated as in Figure  ~\ref{Ue}.

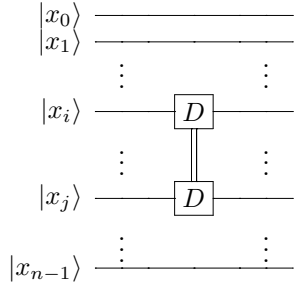
\begin{figure}[H]
\begin{align*}
\Qcircuit @C=1em @R=1em { 
 \lstick{\ket{x_0}} & \qw  & \qw  & \qw& \qw  & \qw  & \qw \\
 \lstick{\ket{x_1}} & \qw  & \qw  & \qw& \qw  & \qw   & \qw\\ 
  &  \vdots &  &&& \vdots\\ 
\lstick{\ket{x_i}} & \qw & \qw & \gate{D}\cwx[2]& \qw  & \qw  & \qw \\
  &  \vdots &  &&& \vdots\\ 
\lstick{\ket{x_j}} & \qw & \qw & \gate{D}\cwx[-2]& \qw  & \qw  & \qw \\
  &  \vdots &  &&& \vdots\\ 
 \lstick{\ket{x_{n-1}}} & \qw  & \qw  & \qw& \qw  & \qw  & \qw \\ 
}
\end{align*}
\caption{A quantum circuit representing the device $D_{i,j}$. \label{Ue}}
\end{figure}

\end{definition}

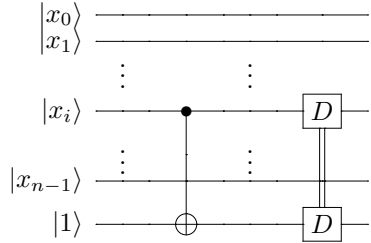
\begin{figure}[H]
\begin{align*}
\Qcircuit @C=1em @R=1em { 
 \lstick{\ket{x_0}} & \qw  & \qw  & \qw& \qw  & \qw  & \qw & \qw  & \qw\\
 \lstick{\ket{x_1}} & \qw  & \qw  & \qw& \qw  & \qw   & \qw & \qw& \qw\\ 
  &  \vdots &  &&& \vdots &  \\ 
\lstick{\ket{x_i}} & \qw & \qw & \ctrl{1}& \qw  & \qw  & \qw & \gate{D}\cwx[2]& \qw\\
   & \vdots &  &&& \vdots & \\ 
 \lstick{\ket{x_{n-1}}} & \qw  & \qw& \qw\qwx & \qw &\qw& \qw  & \qw& \qw \\ 
 \lstick{\ket{1}} &\qw & \qw & \targ\qwx & \qw &\qw & \qw & \gate{D}\cwx[-1]& \qw
}
\end{align*}
\caption{A quantum circuit representing the proposed operator $U_\lambda$.\label{Ul}}
\end{figure}

Our proposed operator $U^i_\lambda$ to check whether a variable $x_i$ is a junta variable or not, will act on a given qubit indexed $i$ and an auxiliary qubit initialized to state $\ket{1}$, as depicted in Figure  \ref{Ul}.

The operator $U_\lambda$, firstly, applies \textit{CNOT} gate on $x_i$ as the control qubit and the auxiliary qubit as a target qubit, and then measures the entanglement between those qubits. It is worth noting that the entanglement will happen if and only if the tested qubit is in a superposition. For further elaboration, let's examine a system $\ket{\varphi}$ of two qubits  one of which is in a superposition. The system can be described as follows:

\begin{equation}
\ket{\varphi}=(\alpha \ket{0}+\beta \ket{1})\otimes \ket{1},
\end{equation}
where $\alpha$ and $\beta$ are complex numbers named the amplitudes and satisfy the relation $\vert \alpha \vert^2+\vert \beta \vert^2=1$. When we apply \textit{CNOT} gate on the system as described earlier, the effect can be illustrated as follows:
\begin{align}
\ket{\varphi^*}&=CNOT \ket{\varphi} \nonumber \\
&= \alpha \ket{01}+\beta \ket{10},
\end{align}
where the second qubit is entangled with the first qubit.
According to Equation ~\eqref{eq:concurrence-ge} and for the given bipartite quantum state $\ket{\varphi^*}$, the concurrence can be restated as follows \cite{walborn,zidan2018novel}:
\begin{equation}
C(\ket{\varphi^*})=\vert 2\alpha\beta\vert.
\end{equation}

\section{A Quantum Algorithm for Testing Variables for the Junta Property}
\label{proposed-algorithm}
\subsection{The Proposed Algorithm}
In this section, we propose a quantum algorithm that tests whether a variable $x_i$ is a junta variable or not, utilizing the entanglement between the tested qubit and an auxiliary qubit. 

\begin{figure}[H]
\begin{align*}
 \Qcircuit @C=1em @R=.7em {
  \lstick{\ket{0}} &  /^n \qw & \gate{H^{\otimes n}} & \multigate{1}{U_f} & \gate{H^{\otimes n}}	&/^n\qw &  \multigate{2}{U^i_\lambda} & \qw \\
  \lstick{\ket{1}} & \qw     & \gate{H}             & \ghost{U_f}  & \gate{H}      & \qw & \ghost{U^i_\lambda} & \qw \\  
 \lstick{\ket{1}} &\qw & \qw & \qw & \qw & \qw & \ghost{U^i_\lambda}  & \qw  \\
 }
\end{align*}
\caption{Quantum circuit for the proposed algorithm.\label{qcircuit-p-algo}}
\end{figure}
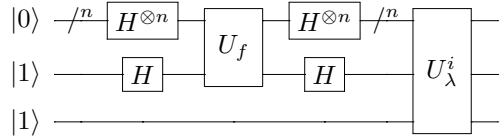

The algorithm is carried quantum mechanically as follows:
\begin{enumerate}

\item \textit{Linearity Check.} Check whether the variable $x_i$ exists in a linear term, as follows:
\label{frst_step}
\begin{enumerate}
\item Prepare a vector $v_0$ of size $n$ initialized with the state $\ket{0}$. 
If $f(v_0)=1$, this will mean that the black-box $U_f$ has a constant term.
\item Prepare a vector $v_1$ of size $n$ initialized with the state $\ket{0}$ except for a $\ket{1}$ at position $i$.
\item If $f(v_1)=0$ and $f(v_0)=1$, or $f(v_1)=1$ and $f(v_0)=0$, this will mean that the black-box $U_f$ has the variable $x_i$ in a linear term. The variable $x_i$ is then considered not a junta variable, and exit.

\end{enumerate}

\item \textit{Register Preparation.}  Prepare a quantum register of $n+2$ qubits. The first $n$ qubits in the state $\ket{0}$ and the extra two qubits in the state $\ket{1}$:
\begin{equation}
\ket{\psi_0}=\ket{0}^{\otimes n}\otimes \ket{1}\otimes \ket{1}.
\end{equation}

\item \textit{Register Initialization.} Apply the Hadamard gate on each qubit of the first $n+1$ qubits to create a uniform superposition of all the possible $N$ states:

\begin{align}
\ket{\psi_1}&=H^{\otimes n+1}\ket{\psi_0} \nonumber\\
&=\frac{1}{\sqrt{N}}\sum_{l=0}^{N-1}\ket{l}\otimes \frac{\ket{0}-\ket{1}}{\sqrt{2}}\otimes \ket{1}.
\end{align}

\item \textit{Apply the oracle $U_f$.} Applying the oracle $U_f$ will mark the solutions satisfying the Boolean function $f$ with a phase shift:

\begin{align}
\ket{\psi_2}&=U_f\ket{\psi_1}\nonumber \\
&=\frac{1}{\sqrt{N}}\Big(\sum_{l=0}^{N-1} \strut^{\prime\prime}\ket{l} - \sum_{l=0}^{N-1}\strut^{\prime}\ket{l}\Big)\otimes \frac{\ket{0}-\ket{1}}{\sqrt{2}}\otimes \ket{1},
\end{align}
where $\sum \strut^{\prime}$ are the states that satisfy the oracle and marked with a phase shift, and $\sum \strut^{\prime\prime}$ otherwise.

\item \textit{Apply Hadamard gates on the Register.} Applying Hadamard gates on the first $n+1$ qubits will put the system in a state that can be described using the language of Fourier analysis as follows:
\begin{equation}
\ket{\psi_3}=\sum_{l=0}^{N-1}\hat{f}(l)\ket{l}\otimes \ket{1}\otimes \ket{1},
\end{equation}
where $\hat{f}(l)$ is the Fourier coefficient of the quantum state $\ket{l}$ and is defined as follows:

\begin{equation}
\hat{f}(l)=\frac{1}{N}\sum_{s=0}^{N-1}(-1)^{f(l)+l\cdot s},
\end{equation}
and $l\cdot s=\sum_{a=0}^{n-1} l_as_a$ is the inner product of the binary strings $l$ and $s$.

\item \textit{Apply the Operator $U^i_\lambda$.} Ignoring the other qubits when applying the oracle $U_\lambda$ on the qubit indexed $i$ and the auxiliary qubit, the subsystem $\ket{\psi_3^i}$ can be described as follows:

\begin{equation}
\ket{\psi_3^i}=\big(\alpha_i\ket{0}+\beta_i\ket{1}\big)\otimes \ket{1},
\end{equation}
and the effect of $U_\lambda$ on $\ket{\psi_3}$  can be viewed for $\ket{\psi_3^i}$ as:


\begin{enumerate}

\item Apply the \textit{CNOT} gate. 
\begin{align}
\ket{\psi_4^i}&=\textit{CNOT}(\alpha_i\ket{0}+\beta_i\ket{1}\big)\otimes \ket{1}\nonumber\\
&=\alpha_i\ket{0,1\oplus\textit{CNOT}(0)}+\beta_i\ket{1, 1\oplus\textit{CNOT}(1)}.
\label{general-system}
\end{align}	

\item Measure the entanglement between the qubit indexed $i$ and the auxiliary qubit.
If there is no entanglement measured, then the variable $x_i$ is considered a junta variable, otherwise the variable $x_i$ is not junta variable.

\end{enumerate}

\end{enumerate}

\subsection{Analysis of the Proposed Algorithm}
\label{analysis}
In this section, we discuss the proposed algorithm with the suggested operator introduced in Section  ~\ref{proposed-operator}. We will analyze the proposed algorithm assuming that the given oracle is a black-box.

In \cite{yang}, Yang \textit{et al.} was able to formulate  an expression of the influence of the variable $x_i$ on the Boolean function $f$ using Bernstein-Vazirani algorithm:

\begin{equation}
I_f(x_i)=\frac{\vert \mathcal{\nu}_1\vert}{N},
\end{equation}
where $\mathcal{\nu}_1$ is the set of all states $x\in \Lambda$ that satisfies the relation \cite{yang}:
\begin{equation}
 x\in \{0,1\}^n \vert f(x\oplus \tau)+f(x)=1,
\end{equation}
and $\tau$ is a binary string of $0$s except for a $1$ at position $i$.

We can say as well that 
\begin{equation}
\vert \mathcal{\nu}_0\vert + \vert \mathcal{\nu}_1\vert=N,
\label{eqn:sum}
\end{equation} 
where $\mathcal{\nu}_0$ is the set of all states $x\in \{0,1\}^n$ that satisfies the relation \cite{yang}:
\begin{equation}
 x\in \{0,1\}^n \vert f(x\oplus \tau)+f(x)\neq 1.
\end{equation}

From Equation ~\eqref{eqn:sum}, we have
\begin{equation}
\frac{\vert \mathcal{\nu}_0\vert}{N}+\frac{\vert \mathcal{\nu}_1\vert}{N}=1,
\end{equation}
and for a single qubit indexed $i$ in a quantum system $\ket{\psi}$, we can state the following:
\begin{equation}
\ket{\psi^i}=\alpha_i\ket{0}+\beta_i\ket{1},
\end{equation}
where $\vert \alpha_i\vert^2=\vert \mathcal{\nu}_0\vert/N$ and $\vert \beta_i\vert^2=\vert \mathcal{\nu}_1\vert/N$.

\paragraph{In case the function $f$ is independent of the variable $x_i$}
When the function $f$ is independent from the variable $x_i$, \textit{i.e.} the variable $x_i$ is a junta variable, then the variable $x_i$ will have no influence on the function $f$, \textit{i.e.} $\vert \mathcal{\nu}_1\vert=0$, thus the subsystem described in Equation ~\eqref{general-system} will be:
\begin{align}
\ket{\psi_4^i}=\ket{01},
\end{align}
and will produce no measurable entanglement between the indicated qubits, \textit{i.e.} $C(\ket{\psi_4^i})=0$.

\paragraph{In case the function $f$ is dependent on the variable $x_i$}
When the function $f$ is dependent on the variable $x_i$, \textit{i.e.}  $x_i$ is not junta, then the variable $x_i$ will have an influence $I_f(x_i)\neq 0$ on the function $f$, and the subsystem in Equation ~\eqref{general-system} can be described as follows:
\begin{align}
\ket{\psi_4^i}&=\alpha_i\ket{01}+\beta_i\ket{10},
\end{align}
and will produce a measurable entanglement between the indicated qubits. The concurrence between those qubits can be defined as follows:
\begin{equation}
C(\ket{\psi_4^i})=\frac{2\times \sqrt{ \vert\mathcal{\nu}_0\vert\times\vert\mathcal{\nu}_1\vert}}{N}.
\end{equation}

In the case that the variable $x_i$ exists in the function $f$ but only in a linear term, the final state of the subsystem will be $\ket{\psi_4^i}=\ket{10}$ with no measurable entanglement. This case can be handled by detecting whether or not the variable $x_i$ exists as a linear term in the function $f$, as covered in Step ~\ref{frst_step} of the proposed algorithm.


\section{Applications of the Proposed Operator for Learning Boolean Functions}
\label{learn}

\subsection{Finding Variables within the Same Terms}

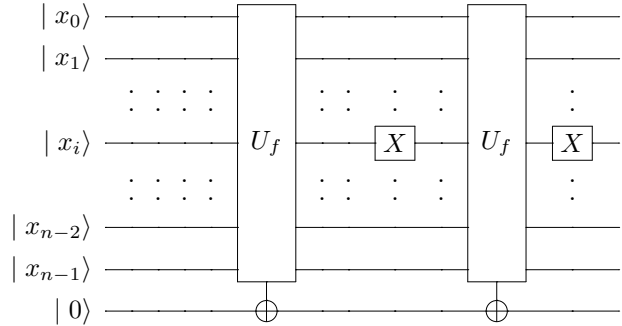
\begin{figure}[H]
\begin{align*}
\Qcircuit @C=1em @R=.7em  {
\lstick{\mid x_0 \rangle}& \qw & \qw & \qw & \qw & \multigate{8}{U_f} & \qw  & \qw & \qw & \qw & \multigate{8}{U_f} & \qw & \qw  \\
\lstick{\mid x_1 \rangle}& \qw & \qw & \qw & \qw & \ghost{U_f}& \qw  & \qw & \qw & \qw & \ghost{U_f} & \qw & \qw\\
& . & . & . & . &  & .  & . & . & . &   & . &   \\
& . & . & . & . &  & .  & . & . & . &   & . &   \\
\lstick{\mid x_i \rangle} & \qw & \qw & \qw & \qw & \ghost{U_f}& \qw  & \qw & \gate{X} & \qw & \ghost{U_f} & \gate{X} & \qw\\
& . & . & . & . &  & .  & . & . & . &   & . &   \\
& . & . & . & . &  & .  & . & . & . &   & . &   \\
\lstick{\mid x_{n-2} \rangle}& \qw & \qw & \qw & \qw & \ghost{U_f}& \qw  & \qw & \qw & \qw & \ghost{U_f} & \qw & \qw\\
\lstick{\mid x_{n-1} \rangle}& \qw & \qw & \qw & \qw & \ghost{U_f}& \qw  & \qw & \qw & \qw & \ghost{U_f} & \qw & \qw\\
\lstick{\mid 0 \rangle}& \qw & \qw & \qw & \qw &  \targ{-1} \qwx & \qw  & \qw & \qw & \qw & \targ \qwx & \qw & \qw \\
}
\end{align*}
\caption{ A quantum circuit for the constructed oracle $U_g$ \cite{elwazan}.\label{Ug}}
\end{figure}

In \cite{elwazan}, we constructed an oracle $U_g$ using two copies of a given black-box $U_f$, and acting on $n+1$ qubits. The oracle $U_g$ is defined as follows:
\begin{equation}
U_g=U_{f_{\bar{x}_i}} U_f.
\end{equation}
An illustration of this circuit is shown in Figure  ~\ref{Ug} .

The oracle $U_g$ has a unique property:
$U_g=I_n$, if and only if $x_i$ is a junta variable, where $I_n$ is the identity matrix of size $2^n\times 2^n$. The effect of the constructed oracle $U_g$ can be illustrated as follows. For any general Boolean function $f$ of $n$ input variables, the function $f$ can be generally defined in terms of a variable $x_i$ as follows \cite{elwazan}:
\begin{align}
f(x_0,\cdots,x_i,\cdots,x_{n-1})=f_{+x_i}\oplus f_{-x_i},
\label{ffx}
\end{align}
such that $f_{+x_i}$ are the terms which contain the variable $x_i$, and $f_{-x_i}$ are the terms which do not contain the variable $x_i$. We define the function $g$ such that 
\begin{equation}
g=f \oplus f_{\bar{x}_i},
\end{equation}
where $f_{\bar{x}_i}=f(x_0,\cdots,\bar{x}_i,\cdots,x_{n-1})$ and $f$ as in Equation~\eqref{ffx}.

The function $g$ can be defined as Equation ~\eqref{ffx}, as follows \cite{elwazan}:
\begin{equation}
g=f_{+x_i}\oplus f_{+\bar{x}_i},
\end{equation}
such that $f_{+\bar{x}_i}$ are the terms which had the variable $x_i$ and are decomposed to lower order terms that do not have the variable $x_i$.

In this section, we will propose an algorithm to find the variables that exist in the same terms with the variable $x_i$, if there is any. This algorithm will output a set $\mathcal{S}$ that contains the indexes of those variables. This proposed application can be used to learn single-term Boolean functions. 
Consider the proposed algorithm in Section  ~\ref{proposed-algorithm} as a subroutine $\textit{PA}(Oracle \textit{ U}, Index \textit{ i})$ that takes an oracle $U$ and the index $i$ of the desired variable for the junta test and  returns $true$ if and only if $x_i$ is junta.
The steps of the proposed algorithm are as follows:

\begin{enumerate}
\item If $\textit{PA}(U_f,i)= true$, \textbf{exit}.
\item Create a set $\mathcal{S}=\{\}$. 
\item Construct the oracle $U_{g}$ for the variable $x_i$.
\item Set $t=0$.
\item $Q=\textit{PA}(U_g,t)$. \label{step}
\item If $Q\neq true$, add $t$ to the set $\mathcal{S}$.
\item If $t<n$,  $t=t+1$ and go to step \ref{step}.
\end{enumerate}

\subsection{Boolean Function Categorization}
A balanced Boolean function is a function which outputs as
many $1$s as $0$s over its possible input $\Lambda$. Meanwhile, a constant Boolean function is such a function which always yield either $1$ or $0$, \textit{i.e.} $\forall x\in\Lambda,~ f:x\rightarrow \{0\}$ or $f:x\rightarrow \{1\}$. Deutsch and Jozsa introduced an exact quantum algorithm \cite{deutsch-jozsa} that solves the problem of identifying whether a given black-box is a balanced Boolean function or a constant Boolean function. We introduce an application of the proposed operator discussed in Section ~\ref{proposed-operator} that discriminates between a constant, a balanced or a Boolean function of other form provided as a black-box, using a variation of Bernstein-Vazirani algorithm. The algorithm is carried quantum mechanically as follows:
\begin{enumerate}
\item \textit{Register Preparation.} Prepare a quantum register of $n+2$ qubits and set the first $n+1$ with the state $\ket{0}$ and the extra qubit with the state $\ket{1}$.

\begin{equation}
\ket{\psi_0}=\ket{0}^{\otimes n+1}\otimes \ket{1}.
\end{equation}

\item \textit{Register Initialization.} Apply Hadamard gates on the first $n$ qubits to get a perfect superposition of all $N$ states.

\begin{align}
\ket{\psi_1}&=H^{\otimes n}\ket{\psi_0} \nonumber \\
&=\frac{1}{\sqrt{N}}\sum_{l=0}^{N-1}\ket{l}\otimes \ket{0}\otimes \ket{1}.
\end{align}

\item \textit{Apply the Oracle $U_f$.} Applying the oracle $U_f$ will yield a quantum system that can be generally expressed as follows:

\begin{align}
\ket{\psi_2}&=U_f\ket{\psi_1}\nonumber \\
&=\frac{1}{\sqrt{N}}\sum_{l=0}^{N-1}\ket{l}\otimes \ket{f(l)}\otimes \ket{1} \nonumber \\
&=\Big(\frac{1}{\sqrt{N}} \sum_{l=0}^{N-1}\strut^{\prime\prime} \ket{l}\otimes \ket{0}+\frac{1}{\sqrt{N}}\sum_{l=0}^{N-1}\strut^{\prime} \ket{l}\otimes \ket{1}\Big)\otimes \ket{1} 
\end{align}
where $\sum^{\prime}$ represents the $M$ states that satisfy the  oracle $U_f$ marked with $\ket{1}$, and $\sum^{\prime\prime}$ otherwise.

\item \textit{Apply Hadamard gates on the Register.} Applying Hadamard gates on the first $n$ qubits of the system will transform the system to the following:

\begin{align}
\ket{\psi_3}&= H^{\otimes n}\ket{\psi_2} \nonumber \\
&=\Big(\frac{1}{\sqrt{N}} \sum_{l=0}^{N-1}\strut^{\prime\prime} H\ket{l}\otimes \ket{0}+\frac{1}{\sqrt{N}}\sum_{l=0}^{N-1}\strut^{\prime} H\ket{l}\otimes \ket{1}\Big)\otimes \ket{1} \nonumber \\
&=\Big(\sum_{l=0}^{N-1}\tilde{f}_0(l)\ket{l}\otimes \ket{0} + \sum_{l=0}^{N-1}\tilde{f}_1(l)\ket{l}\otimes \ket{1}\Big)\otimes \ket{1},
\end{align} 
where $\tilde{f}_0$ and $\tilde{f}_1$ are 
\begin{align}
\tilde{f}_0(l)&=\frac{1}{N}\sum_{s=0}^{N-1}\strut^{\prime\prime}(-1)^{l\cdot s}, \\
\tilde{f}_1(l)&=\frac{1}{N}\sum_{s=0}^{N-1}\strut^{\prime}(-1)^{l\cdot s}.
\end{align}

\item \textit{Apply the Operator $U^n_\lambda$.} Applying $U^n_\lambda$ on the quantum system 
will yield a subsystem:
\begin{equation}
\ket{\psi^n_4}=\alpha_n\ket{01}+\beta_n\ket{10},
\end{equation}
and the measured concurrence $C(\ket{\psi^n_4})$ can be expressed as follows:
\begin{equation}
C(\ket{\psi^n_4})=2\times \frac{\sqrt{M(N-M)}}{N}.
\label{eq:M}
\end{equation}


If $C=0$, then the given oracle represents a \textit{constant} Boolean function, and if $C=1/2$ then the oracle represents a \textit{balanced} Boolean function, but otherwise this will mean that the black-box represents a Boolean function of other form.

\end{enumerate}

One of the potential applications of the proposed Boolean function categorization algorithm is that it can be utilized to know the number of solutions $M$ to any given black-box by solving Equation~\eqref{eq:M} for unknown $M$.

\section{Conclusion}
\label{conclusion}
In this paper, we proposed a fast quantum algorithm that uses entanglement to test the junta property of a certain variable in a Boolean function given as a black-box. We transformed the junta variable testing to measuring entanglement between the variable being tested and an auxiliary qubit. 

The algorithm creates entanglement between the tested variable and the auxiliary qubit if the variable being tested is not a junta variable. If the variable being tested is a junta variable, there will be no measurable entanglement using concurrence measure, otherwise the variable is considered not junta.  Testing whether a Boolean function is a $k$-junta function or not will require $\mathcal{O}(n)$ oracle calls, for unknown $k$.


The paper also introduced applications for learning black-boxes. The first application finds the variables in the same product terms with the variable under test using $\mathcal{O}(n)$ oracle calls; this application can be utilized to learn single-term Boolean functions. The second application is to categorize any given black-box representing an unknown Boolean function to either a constant, a balanced or a Boolean function of other form.

\section*{Acknowledgment}
We would like to thank Prof. Mahmoud Abdel-Aty, Dr. Abdel-Haleem Abdel-Aty and Mohammed Zidan for their valuable suggestions on entanglement concurrence measure.

\bibliographystyle{IEEEtran}
\bibliography{references1}

\begin{thebibliography}{10}
\providecommand{\url}[1]{#1}
\csname url@samestyle\endcsname
\providecommand{\newblock}{\relax}
\providecommand{\bibinfo}[2]{#2}
\providecommand{\BIBentrySTDinterwordspacing}{\spaceskip=0pt\relax}
\providecommand{\BIBentryALTinterwordstretchfactor}{4}
\providecommand{\BIBentryALTinterwordspacing}{\spaceskip=\fontdimen2\font plus
\BIBentryALTinterwordstretchfactor\fontdimen3\font minus
  \fontdimen4\font\relax}
\providecommand{\BIBforeignlanguage}[2]{{%
\expandafter\ifx\csname l@#1\endcsname\relax
\typeout{** WARNING: IEEEtran.bst: No hyphenation pattern has been}%
\typeout{** loaded for the language `#1'. Using the pattern for}%
\typeout{** the default language instead.}%
\else
\language=\csname l@#1\endcsname
\fi
#2}}
\providecommand{\BIBdecl}{\relax}
\BIBdecl

\bibitem{Feynman1986}
R.~P. Feynman, ``Quantum mechanical computers,'' \emph{Foundations of physics},
  vol.~16, no.~6, pp. 507--531, 1986.

\bibitem{deutsch}
D.~Deutsch, ``Quantum theory, the church--turing principle and the universal
  quantum computer,'' \emph{Proceedings of the Royal Society A}, vol. 400, no.
  1818, pp. 97--117, 1985.

\bibitem{loyed}
S.~Lloyd, ``{A potentially realizable quantum computer.}'' \emph{Science}, vol.
  261, no. 5128, pp. 1569--1571, 1993.

\bibitem{buhrman}
H.~Buhrman, R.~Cleve, and A.~Wigderson, ``Quantum vs. classical communication
  and computation,'' in \emph{Proceedings of the thirtieth annual ACM symposium
  on Theory of computing}.\hskip 1em plus 0.5em minus 0.4em\relax ACM, 1998,
  pp. 63--68.

\bibitem{grover}
L.~K. Grover, ``Quantum mechanics helps in searching for a needle in a
  haystack,'' \emph{Physical review letters}, vol.~79, no.~2, p. 325, 1997.

\bibitem{grover-optimal}
C.~Zalka, ``{Grover's quantum searching algorithm is optimal},'' \emph{Physical
  Review A}, vol.~60, no.~4, pp. 2746--2751, 1999.

\bibitem{boyer}
M.~Boyer, G.~Brassard, P.~H{\o}yer, and A.~Tapp, ``Tight bounds on quantum
  searching,'' \emph{Fortschritte der Physik: Progress of Physics}, vol.~46,
  no. 4-5, pp. 493--505, 1998.

\bibitem{shor}
P.~Shor, ``{Polynomial-time algorithms for prime factorization and discrete
  logarithms on a quantum computer},'' \emph{SIAM Journal on Computing},
  vol.~26, no.~5, pp. 1484--1509, 1997.

\bibitem{EPR}
A.~Einstein, B.~Podolsky, and N.~Rosen, ``Can quantum-mechanical description of
  physical reality be considered complete?'' \emph{Physical Review}, vol.~47,
  no.~10, p. 777, 1935.

\bibitem{neilson-chuang}
M.~A. Nielsen and I.~L. Chuang, \emph{{Quantum computation and quantum
  information}}.\hskip 1em plus 0.5em minus 0.4em\relax AAPT, 2010.

\bibitem{younes-miller}
A.~Younes, J.~Rowe, and J.~Miller, ``{Enhanced quantum searching via
  entanglement and partial diffusion},'' \emph{Physica D.}, vol. 237, no.~8,
  pp. 1074--1078, 2008.

\bibitem{younes}
A.~Younes, ``Constant-time quantum algorithm for the unstructured search
  problem,'' \emph{arXiv preprint arXiv:0811.4247}, 2008.

\bibitem{quantum-satellite}
C.~J. Pugh, S.~Kaiser, J.-P. Bourgoin, J.~Jin, N.~Sultana, S.~Agne,
  E.~Anisimova, V.~Makarov, E.~Choi, B.~L. Higgins \emph{et~al.}, ``Airborne
  demonstration of a quantum key distribution receiver payload,'' \emph{Quantum
  Science and Technology}, vol.~2, no.~2, p. 024009, 2017.

\bibitem{quantum-satellite-exp}
S.-k. Liao, W.-q. Cai, W.-y. Liu, L.~Zhang, Y.~Li, J.-g. Ren, J.~Yin, Q.~Shen,
  Y.~Cao, Z.-P. Li, F.-Z. Li, X.-W. Chen, L.-H. Sun, J.-J. Jia, J.-C. Wu, X.-J.
  Jiang, J.-F. Wang, Y.-M. Huang, Q.~Wang, Y.-L. Zhou, L.~Deng, T.~Xi, L.~Ma,
  T.~Hu, Q.~Zhang, Y.-A. Chen, N.-L. Liu, X.-B. Wang, Z.-C. Zhu, C.-Y. Lu,
  R.~Shu, C.-Z. Peng, J.-Y. Wang, and J.-W. Pan, ``{Satellite-to-ground quantum
  key distribution},'' \emph{Nature}, 2017.

\bibitem{quantum-internet}
Q.-C. Sun, Y.-L. Mao, S.-J. Chen, W.~Zhang, Y.-F. Jiang, Y.-B. Zhang, W.-J.
  Zhang, S.~Miki, T.~Yamashita, H.~Terai, X.~Jiang, T.-Y. Chen, L.-X. You,
  X.-F. Chen, Z.~Wang, J.-Y. Fan, Q.~Zhang, and J.-W. Pan, ``{Quantum
  teleportation with independent sources and prior entanglement distribution
  over a network},'' \emph{Nature Photonics}, no.~10, pp. 671--675, 2016.

\bibitem{cloud}
J.~Hald, J.~S{\o}rensen, C.~Schori, and E.~Polzik, ``Spin squeezed atoms: a
  macroscopic entangled ensemble created by light,'' \emph{Physical Review
  Letters}, vol.~83, no.~7, p. 1319, 1999.

\bibitem{ion-trapped}
H.~H{\"a}ffner, W.~H{\"a}nsel, C.~Roos, J.~Benhelm, M.~Chwalla, T.~K{\"o}rber,
  U.~Rapol, M.~Riebe, P.~Schmidt, C.~Becher \emph{et~al.}, ``Scalable
  multiparticle entanglement of trapped ions,'' \emph{Nature}, vol. 438, no.
  7068, p. 643, 2005.

\bibitem{photons}
C.-Y. Lu, X.-Q. Zhou, O.~G{\"{u}}hne, W.-B. Gao, J.~Zhang, Z.-S. Yuan,
  A.~Goebel, T.~Yang, and J.-W. Pan, ``{Experimental entanglement of six
  photons in graph states},'' \emph{Nature Physics}, vol.~3, no.~2, pp. 91--95,
  2007.

\bibitem{li-zhao}
M.~Li, M.~J. Zhao, S.~M. Fei, and Z.~X. Wang, ``{Experimental detection of
  quantum entanglement},'' \emph{Frontiers of Physics}, vol.~8, no.~4, pp.
  357--374, 2013.

\bibitem{zhou}
L.~Zhou and Y.~B. Sheng, ``{Concurrence measurement for the two-qubit optical
  and atomic states},'' \emph{Entropy}, vol.~17, no.~6, pp. 4293--4322, 2015.

\bibitem{zhang}
L.~H. Zhang, M.~Yang, and Z.~L. Cao, ``{Direct measurement of the concurrence
  for two-photon polarization entangled pure states by parity-check
  measurements},'' \emph{Physical Review A}, vol. 377, no. 21-22, pp.
  1421--1424, 2013.

\bibitem{regula-adesso}
B.~Regula and G.~Adesso, ``{Entanglement quantification made easy: polynomial
  measures invariant under convex decomposition},'' \emph{Physical Review
  Letters}, vol. 116, no.~7, pp. 1--5, 2016.

\bibitem{e-detection}
O.~G{\"u}hne and G.~T{\'o}th, ``Entanglement detection,'' \emph{Physics
  Reports}, vol. 474, no. 1-6, pp. 1--75, 2009.

\bibitem{makinsey-report}
{McKinsey {\&} Company}, ``{Big data: The next frontier for innovation,
  competition, and productivity},'' \emph{McKinsey Global Institute}, no. June,
  p. 156, 2011.

\bibitem{survey}
A.~Montanaro and R.~de~Wolf, ``A survey of quantum property testing,''
  \emph{arXiv preprint arXiv:1310.2035}, 2013.

\bibitem{blum1}
A.~Blum, ``Relevant examples and relevant features: Thoughts from computational
  learning theory,'' in \emph{AAAI Fall Symposium on ‘Relevance}, vol.~5,
  1994, p.~1.

\bibitem{blum2}
A.~L. Blum and P.~Langley, ``Selection of relevant features and examples in
  machine learning,'' \emph{Artificial intelligence}, vol.~97, no. 1-2, pp.
  245--271, 1997.

\bibitem{atici-servedo}
A.~At{\i}c{\i} and R.~A. Servedio, ``Quantum algorithms for learning and
  testing juntas,'' \emph{Quantum Information Processing}, vol.~6, no.~5, pp.
  323--348, 2007.

\bibitem{Bernstein-Vazirani}
E.~Bernstein and U.~Vazirani, ``Quantum complexity theory,'' \emph{SIAM Journal
  on computing}, vol.~26, no.~5, pp. 1411--1473, 1997.

\bibitem{yang}
H.~Li and L.~Yang, ``{A quantum algorithm for approximating the influences of
  Boolean functions and its applications},'' \emph{Quantum Information
  Procesing}, vol.~14, no.~6, pp. 1787--1797, 2015.

\bibitem{Ambainis}
A.~Ambainis, A.~Belovs, O.~Regev, and R.~De~Wolf, ``Efficient quantum
  algorithms for (gapped) group testing and junta testing,'' in
  \emph{Proceedings of the twenty-seventh annual ACM-SIAM symposium on Discrete
  algorithms}.\hskip 1em plus 0.5em minus 0.4em\relax Society for Industrial
  and Applied Mathematics, 2016, pp. 903--922.

\bibitem{Sterrett}
A.~Sterrett, ``On the detection of defective members of large populations,''
  \emph{The Annals of Mathematical Statistics}, vol.~28, no.~4, pp. 1033--1036,
  1957.

\bibitem{elwazan}
K.~El-Wazan, A.~Younes, and S.~Doma, ``A quantum algorithm for testing juntas
  in boolean functions,'' \emph{arXiv preprint arXiv:1701.02143}, 2017.

\bibitem{reed-muller}
A.~Younes and J.~Miller, ``{Representation of Boolean quantum circuits as
  Reed-Muller expansions},'' \emph{International Journal of Electronics},
  vol.~91, pp. 1--12, 2003.

\bibitem{hill}
S.~Hill and W.~K. Wootters, ``{Entanglement of a pair of quantum bits},''
  \emph{Physical Review Letters}, vol.~78, no.~26, pp. 5022--5025, 1997.

\bibitem{walborn}
S.~P. Walborn, P.~H. {Souto Ribeiro}, L.~Davidovich, F.~Mintert, and
  A.~Buchleitner, ``{Experimental determination of entanglement with a single
  measurement},'' \emph{Nature}, vol. 440, no. 7087, pp. 1022--1024, 2006.

\bibitem{zidan2018novel}
M.~Zidan, A.-H. Abdel-Aty, A.~Younes, E.~Zanaty, I.~El-khayat, and
  M.~Abdel-Aty, ``A novel algorithm based on entanglement measurement for
  improving speed of quantum algorithms,'' \emph{Applied Mathematics \&
  Information Sciences}, vol.~12, no.~1, pp. 265--269, 2018.

\bibitem{deutsch-jozsa}
D.~Deutsch and R.~Jozsa, ``Rapid solution of problems by quantum computation,''
  \emph{Proceedings of the Royal Society A}, vol. 439, no. 1907, pp. 553--558,
  1992.

\end{thebibliography}

\end{document}